\documentclass[pra,twocolumn,superscriptaddress,showpacs,a4paper]{revtex4}

\usepackage[pdftex]{graphicx}
\usepackage{amssymb}
\usepackage{amsmath}
\usepackage{epsfig}
\usepackage{latexsym}
\usepackage{color}
\usepackage{rotating}

\begin{document}

\title{Nanomechanical squeezing with detection via a microwave cavity}

\author{M. J. Woolley} 
\affiliation{Department of Physics, School of Physical Sciences, University
of Queensland, St Lucia, QLD 4072, Australia}
\author{A. C. Doherty}
\affiliation{Department of Physics, School of Physical Sciences, University
of Queensland, St Lucia, QLD 4072, Australia}
\author{G. J. Milburn}
\affiliation{Department of Physics, School of Physical Sciences, University
of Queensland, St Lucia, QLD 4072, Australia}
\author{K. C. Schwab}
\affiliation{Department of Physics, Cornell University, Ithaca, NY 14853, USA}

\begin{abstract}
We study a parametrically-driven nanomechanical resonator capacitively coupled to a microwave cavity. If the nanoresonator can be cooled to near its quantum ground state then quantum squeezing of a quadrature of the nanoresonator motion becomes feasible. We consider the adiabatic limit in which the cavity mode is slaved to the nanoresonator mode. By driving the cavity on its red-detuned sideband, the squeezing can be coupled into the microwave field at the cavity resonance. The red-detuned sideband drive is also compatible with the goal of ground state cooling. Squeezing of the output microwave field may be inferred using a technique similar to that used to infer squeezing of the field produced by a Josephson parametric amplifier, and subsequently, squeezing of the nanoresonator motion may be inferred. We have calculated the output field microwave squeezing spectra and related this to squeezing of the nanoresonator motion, both at zero and finite temperature. Driving the cavity on the blue-detuned sideband, and on both the blue and red sidebands, have also been considered within the same formalism.  
\end{abstract}


\maketitle

\section{INTRODUCTION}

Quantum squeezing of mechanical motion is but one of a number of inherently quantum phenomena that may soon be observed in macroscopic mechanical systems \cite{Blencowe1}. Squeezing of mechanical motion was first demonstrated in the context of (classical) thermomechanical noise squeezing \cite{Rugar}. A micromechanical cantilever was driven parametrically via a coupled capacitor plate, and the motion of the cantilever was detected via a fiber-optic sensor. The thermal noise of the cantilever in one quadrature of the motion was observed to be reduced below the equilibrium value. It has been shown that \emph{quantum} squeezing \cite{Blencowe} of the motion of a nanomechanical resonator, analogous to the single-mode squeezing of light in a degenerate parametric amplifier (DPA) below threshold \cite{Wu}, is feasible using a similar approach. However, the detection of such states is problematic. 

A number of schemes have been proposed for generating squeezed states of nanoresonators. Coupling to a Cooper pair box charge qubit \cite{Rabl} or to a SQUID \cite{Zhou} have both been suggested, though here an additional qubit read-out device is required, the proposed systems are relatively complicated and cooling of a nanoresonator is yet to be demonstrated in these systems. A scheme based on circuit QED has also been proposed \cite{Huo}, though here the measurement scheme has not been made explicit. In a more theoretical work, Jacobs \cite{Jacobs} has shown how to generate a range of nonlinear Hamiltonians for a nanoresonator via control of a coupled qubit. Other proposals are based on quantum non-demolition measurement and feedback \cite{Wiseman1}, via either a coupled single electron transistor \cite{Ruskov} or coupled microwave cavity \cite{Clerk}. However, these schemes require truly quantum-limited measurement in order to achieve the required feedback and squeezing. Our scheme facilitates the generation and detection of a squeezed mechanical state without detection at the single photon level, a task that would be feasible in optomechanical systems but is not feasible in electromechanical systems. 

\begin{figure}[th]
\includegraphics[scale=0.5]{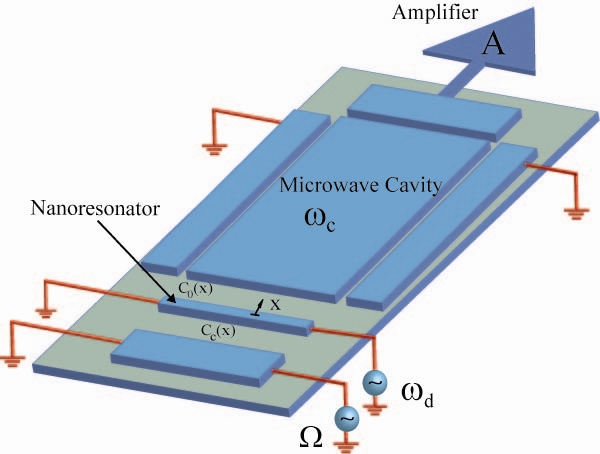}
\caption{Schematic of a nanoresonator capacitively coupled to a microwave cavity in the form of a superconducting coplanar waveguide. The nanoresonator is driven parametrically, and hence its motion may be squeezed, via a coupled capacitor plate. By driving the cavity on its red sideband, detuned by the mechanical resonance, the squeezing may be coupled into squeezing of the microwave field. This squeezing may then be inferred by amplification and homodyne detection of the output microwave field. } 
\label{fig1}
\end{figure}

In our proposal, the nanoresonator is driven parametrically via a coupled capacitor plate. A coupled microwave cavity, in the form of a superconducting coplanar waveguide, functions as a transducer for the nanoresonator motion. A schematic is shown in Fig.~\ref{fig1}. By driving the cavity on its red-detuned sideband corresponding to the mechanical resonance frequency, in the adiabatic limit in which the cavity mode is slaved to the nanoresonator mode, squeezing of the nanoresonator may be coupled into squeezing of the microwave field. Significantly, this driving condition is compatible with the goal of ground-state cooling of the nanoresonator. A quadrature of the output microwave field is measured by homodyne detection after amplification and one may infer quantum squeezing of the field, and hence of the nanoresonator motion, in a manner similar to that used for the Josephson parametric amplifier \cite{CB1}.   

Recently, the motion of a nanoresonator coupled to a microwave cavity has been detected \cite{Lehnert2} and cooling of the nanoresonator has been demonstrated \cite{Lehnert3,Lehnert4}. It is believed that this sideband cooling should be able to cool nanoresonators to near their quantum ground state \cite{Girvin1,Kippenberg1}, and hence these advances open up the possibility for the observation of quantum squeezing. 

In Sec.~II we describe the system and derive its Hamiltonian. Sec.~III provides a general description of the system dynamics using the input-output formalism of quantum optics \cite{Collett}. We derive effective Hamiltonians for the cases where the cavity is driven on its sidebands corresponding to the mechanical resonance frequency. The steady-state squeezing of the nanoresonator quadratures is calculated in Sec.~IV. We justify our assertion that the cavity can function as a near quantum-limited transducer in Sec.~V by adiabatically eliminating the cavity mode from the total effective master equation. Sec.~VI contains the calculation of the output microwave field squeezing spectra, and these spectra are related to the squeezing of the nanoresonator quadratures. The zero and finite temperature mechanical bath cases are studied, and an assessment of experimental feasibility is provided. Our conclusions are presented in Sec.~VII. 

\section{SYSTEM AND HAMILTONIAN}
Suppose the cavity has a resonance frequency $\omega_c$, and the nanoresonator has a resonance frequency $\nu$ and mass $m$. The uncoupled cavity may be described by a lumped parameter circuit with an equivalent inductance $L$ and equivalent capacitance $C$. The capacitive coupling between the cavity and nanoresonator, as a function of the nanoresonator displacement from its equilibrium position $x$, is added in parallel to the equivalent capacitor and may be approximated by $ C_0(x)=C_0\left(1-x/d\right)$ where $C_0$ represents an equilibrium capacitance and $d$ is the equilibrium nanoresonator-cavity separation. Thus the coupled cavity has an equivalent capacitance $C_\Sigma = C + C_0$, such that the coupled resonance frequency is $\omega_c = 1/\sqrt{LC_\Sigma}$. The capacitive energy of the system is then $ Q^2/2C_\Sigma + (\beta/2dC_\Sigma)xQ^2 $ where $\beta = C_0/C_\Sigma$. 

Parametric driving of the nanoresonator at frequency $\Omega$ with strength $\chi$ corresponds to a modulation of the nanoresonator's effective spring constant $k(t) = 4m\nu \left( \chi^*e^{i\Omega t}+\chi e^{-i\Omega t}\right)$. Such a drive may be realized via a capacitive coupling $C_c(x)$, with $ k(t) = \partial^2/\partial x^2\left[C_c(x)V^2(t)/2 \right]$ where $ C_c(x) = C_{c0}\left( 1 - x(t)/x_{c0} + x^2(t)/2x^2_{c0} - ...  \right)$ and $ V(t) = V_0 + V_P \cos \Omega t$. Considering only the component at $\Omega $, we find $ k(t) = \left( C_{c0} V_0V_P\cos \Omega t \right)/ x^2_{c0}$.

The cavity is driven at the frequency (or frequencies) $\omega_{d1}$ (and $\omega_{d2}$), according to the electric potential $e(t)=2\sqrt{2\hbar\omega_c L}\mathcal{E}_1 (e^{i\omega_{d1} t}+e^{-i\omega_{d1} t}) + 2\sqrt{2\hbar\omega_c L}\mathcal{E}_2 (e^{i\omega_{d2} t}+e^{-i\omega_{d2} t}) $. Thus the system is described by the classical Hamiltonian, 
\begin{eqnarray}
H&=&\frac{p^2}{2m} + \frac{1}{2}m\nu^2 x^2 + \frac{\Phi^2}{2L} + \frac{Q^2}{2C_\Sigma} \nonumber \\ 
& & \ + \frac{\beta}{2dC_\Sigma}xQ^2 +\frac{1}{2}e(t)Q + \frac{1}{2}k(t)x^2,  
\end{eqnarray}
where $(x,p)$ are the canonical position and momentum coordinates of the nanoresonator, $(Q,\Phi)$ are the canonical coordinates for the cavity (representing the charge on the equivalent capacitor $C_\Sigma$ and flux through the equivalent inductor $L$). 

Now we may quantize the Hamiltonian by imposing the commutation relations $[\hat{x},\hat{p}]=i\hbar$ and $[\hat{Q},\hat{\Phi}]=i\hbar$. In the Schr\"{o}dinger picture, 
\begin{eqnarray}
H_S&=&\hbar\omega_c a^\dagger a+\hbar\nu b^\dagger b+ \frac{1}{2}\hbar \kappa (b+b^\dagger)(a+a^\dagger)^2 \nonumber \\  
& &\ +\hbar(\chi^*e^{i\Omega t}+\chi e^{-i\Omega  t})(b+b^\dagger)^2 \nonumber \\
& &\ \ \ +\hbar (\mathcal{E}^*_1 e^{i\omega_{d1} t}+\mathcal{E}_1e^{-i\omega_{d1} t})(a+a^\dagger) \nonumber \\
& &\ \ \ \ \ +\hbar (\mathcal{E}^*_2 e^{i\omega_{d2} t}+\mathcal{E}_2 e^{-i\omega_{d2} t})(a+a^\dagger), 
\end{eqnarray}
where the cavity and nanoresonator operators are
\begin{subequations}
\begin{eqnarray}
a & = & \sqrt{\frac{\omega_cL}{2\hbar}} \hat{Q}+ \frac{i}{\sqrt{2\hbar \omega_cL}}\hat{\Phi}, \\
b & = & \sqrt{\frac{m\nu}{2\hbar}} \hat{x}+ \frac{i}{\sqrt{2\hbar m\nu}}\hat{p},
\end{eqnarray}
\end{subequations}
respectively, and the coupling constant is
\begin{equation}
\kappa = \frac{\beta\omega_c}{2} \frac{\Delta x}{d}, \label{eq:kappaapprox}
\end{equation}
with $\Delta x=(\hbar/2m\nu)^{1/2}$ being the half-width of the nanoresonator ground state wavefunction. The phases of the drive terms are all retained for generality; locking these drives should be feasible at microwave frequencies.

Suppose that the microwave cavity is driven on \emph{one sideband} $(\mathcal{E}_2 = 0 )$. Then setting $\Omega = 2\nu$ and transforming to an interaction picture with respect to $ H^1_0 = \hbar \omega_{d1} a^\dagger a + \hbar \nu b^\dagger b$, retaining energy-conserving terms, leads to 
\begin{eqnarray}
H^1_I &=& \hbar\delta_1 a^{\dagger}a +\hbar\kappa X\left(t\right)a^\dagger a \nonumber \\
& & \ +\hbar \left( \mathcal{E}^*_1 a+ \mathcal{E}_1a^\dagger\right) + \hbar\left( \chi^*b^2 + \chi b^{\dagger2}\right), \label{Hint}
\end{eqnarray}
where $\delta_1 = \omega_c - \omega_{d1}$ is the detuning between the cavity resonance and the drive, and $ X\left(t\right)=be^{-i\nu t}+b^\dagger e^{i\nu t}$. 

Alternatively, following Clerk \emph{et al.} \cite{Clerk}, suppose the cavity is driven on two sidebands. This allows a back-action evading measurement of one quadrature of the nanoresonator's motion \cite{Braginsky1}. Now transforming to an interaction picture with respect to $H^2_0 = \hbar \omega_c a^\dagger a + \hbar \nu b^\dagger b $, and again setting $\Omega = 2\nu $ and using the rotating wave approximation,
\begin{eqnarray}
H^2_I &=& \hbar\kappa X\left(t\right)a^\dagger a +\hbar (\mathcal{E}^*_1 a e^{-i\delta_1 t} + \mathcal{E}_1 a^\dagger e^{i\delta_1 t} ) \nonumber \\
& & +\hbar (\mathcal{E}^*_2 a e^{-i\delta_2 t} + \mathcal{E}_2 a^\dagger e^{i\delta_2 t} ) + \hbar\left( \chi^*b^2 + \chi b^{\dagger2}\right), \nonumber \label{Hint2} \\
\end{eqnarray}
where $ \delta_2 = \omega_c - \omega_{d2} $. 

The last terms in both (\ref{Hint}) and (\ref{Hint2}) are then exactly that of a DPA below threshold. The terms proportional to $\kappa$ represent a low frequency modulation, at the mechanical resonance frequency, of the cavity resonance frequency. In principle, this interaction could be used to continuously monitor the position of the nanoresonator \cite{Jacobs2}. 

This interaction also writes sidebands onto the cavity transmission spectrum at integer multiples of the mechanical resonance frequency, and we now consider this sideband picture. By driving the cavity on these sidebands, the cavity field on resonance couples to the slowly-varying quadratures of the nanoresonator motion. Monitoring the motion via a sideband of the drive also allows the ``meter'' cavity mode to be initially near its quantum ground state, and, as shall be demonstrated, will provide a controllable nanoresonator-cavity coupling. 

\section{QUANTUM LANGEVIN EQUATIONS}
Damping of the microwave cavity, at a rate $\mu << \omega_c $, is described using a generic quantum optical master equation and the associated quantum Langevin equations \cite{Walls-GJM}. It is assumed that the cavity's internal losses are negligible compared with damping due to out-coupling of the field, such that the cavity is a good transducer. 

It is assumed that the nanoresonator undergoes a quantum Brownian motion. Then it is preferable to use a quantum Langevin approach, rather than a master equation approach. One may use non-Markovian correlation functions to describe the input noise, though in the limit $\nu >> \gamma $ (the mechanical damping rate), the input noise operators become delta-correlated in time \cite{Vitali1} and the use of a quantum optics master equation suffices. It is believed that, at low temperatures, the bath to which the nanoresonator is predominantly coupled may take the form of a spin bath \cite{Mohanty1}; in principle, this may be dealt with via non-Markovian correlation functions, though this approach is not pursued here. 

\subsection{Cavity Driven on One Sideband}
Hence, for the cavity driven on one sideband, the Hamiltonian (\ref{Hint}) leads to the quantum Langevin equations,
\begin{subequations}
\begin{eqnarray}
\dot{a}(t) & = & -i\delta_1 a(t) -i\mathcal{E}_1 -\frac{\mu}{2}a(t) + \sqrt{\mu}a_{in}(t) \nonumber \\
 & & \ \ \ \ \ \ -i\kappa \left[b(t)e^{-i\nu t}+b^\dagger(t) e^{i\nu t}\right]a(t), \label{wholesystem1} \\
\dot{b}(t) & = & -2i\chi b^\dagger(t) -i\kappa a^\dagger(t) a(t) e^{i\nu t} -\frac{\gamma}{2}b(t)+\sqrt{\gamma}b_{in}(t). \label{wholesystem2} \nonumber \\
& & 
\end{eqnarray} 
\end{subequations}
The non-zero input noise correlation functions are
\begin{subequations}
\begin{eqnarray}
\left\langle a^\dagger_{in}\left( t \right), a_{in}\left( t' \right)  \right\rangle & = & \delta\left(t-t' \right), \label{cavnoise} \\
\left\langle b^\dagger_{in}\left( t \right), b_{in}\left( t' \right)  \right\rangle & = & n^0_m \delta\left(t-t' \right) \label{nrnoise}, 
\end{eqnarray}
\end{subequations}
with $n^0_m$ being the thermal occupancy of the mechanical bath mode at the mechanical resonance frequency,
\begin{equation}
n^0_m = \left[ \exp \left(\frac{\hbar \nu}{k T_m}\right)-1\right]^{-1}, 
\end{equation} 
where $T_m$ is the effective mechanical bath temperature, and it is assumed that the cavity resonance is at a sufficiently high frequency that the cavity may be considered to be damped into a zero-temperature bath.

Assuming that we are in the resolved sideband regime,
\begin{equation}
\left|\delta_1 \right|>> \mu, \label{rsb1}
\end{equation}
solutions to (\ref{wholesystem1})-(\ref{wholesystem2}) should be well-approximated by the ansatz
\begin{subequations}
\begin{eqnarray}
a (t) &=& a_0 (t) + a_+ (t)e^{-i\nu t}+ a_- (t)e^{i\nu t}, \label{eq:assumedalpha}\\
b (t) &=& b_0 (t), \label{eq:assumedbeta}
\end{eqnarray}
\end{subequations}
where the subscripts $+$ and $-$ denote sidebands above and below, respectively, the cavity drive frequency. Substituting this into (\ref{wholesystem1})-(\ref{wholesystem2}) and equating frequency components, we obtain 
\begin{widetext}
\begin{subequations}
\begin{eqnarray}
\dot{a}_0(t) &=& -i\delta_1 a_0(t) - i\mathcal{E}_1 -i\kappa\left[ a_+(t)b^\dagger_0(t)+a_-(t)b_0(t)\right] -\frac{\mu}{2} a_0(t) +\sqrt{\mu}a_{o,in}(t) ,  \label{eq:kappagone} \\
\dot{a}_+(t) &=& -i(\delta_1-\nu)a_+(t) -i\kappa a_0(t) b_0(t) -\frac{\mu}{2}a_+(t) +\sqrt{\mu}a_{+,in}(t), \label{eq:qle2}\\
\dot{a}_-(t) &=& -i(\delta_1+\nu) a_-(t)  -i\kappa a_0(t) b^\dagger_0(t) \label{eq:qle3} -\frac{\mu}{2} a_-(t) +\sqrt{\mu}a_{-,in}(t), \\ 
\dot{b}_0(t) &=& -2i\chi b^\dagger_0(t) -i\kappa \left[a^\dagger_0(t) a_+(t) +a_0(t) a^\dagger_-(t)\right] -\frac{\gamma}{2} b_0(t) +\sqrt{\gamma}b_{o,in}(t) . \label{eq:qle4}
\end{eqnarray}
\end{subequations}
\end{widetext}
If the cavity is driven either on the first blue or the first red sideband of its resonance, one sideband of the drive will be resonant with the cavity and the other sideband will be far from resonance. We may then neglect the off-resonant sideband, find the steady-state at the drive frequency, and analyze the reduced system composed of the sideband on resonance and the mechanical element. 

\subsubsection{Cavity Driven on Blue Sideband}
If the cavity is driven on its first blue sideband, 
\begin{equation}
\omega_d = \omega_c + \nu, \ \ \ \ \ \ \ \ \left(i.e. \ \ \  \delta_1 = -\nu \right),
\end{equation}
the oscillation of the blue sideband of the driving field is off-resonance, and we may neglect $a_+ (t)$. Assuming $\kappa << \mu , \left|\delta_1 \right|, \left|\mathcal{E}_1\right|$ and, without loss of generality, that $\mathcal{E}_1 $ is real and positive, then (\ref{eq:kappagone}) gives the steady-state at the blue sideband drive $\left\langle a^b_0(t\rightarrow \infty) \right\rangle =\mathcal{E}_1/\nu$. Then (\ref{eq:qle3}) and (\ref{eq:qle4}), with the corresponding Hermitian conjugate equations and now dropping sideband subscripts, yield  
\begin{subequations}
\begin{eqnarray}
\dot{a}(t) & = & -\frac{\mu}{2}a(t) -igb^\dagger(t) + \sqrt{\mu}a_{in}(t), \label{eq:qle5} \\
\dot{a}^\dagger(t) & = & -\frac{\mu}{2}a^\dagger(t) +igb(t) + \sqrt{\mu}a^\dagger_{in}(t), \\
\dot{b}(t) & = & -\frac{\gamma}{2}b(t) -2i\chi b^\dagger(t) -iga^\dagger(t) + \sqrt{\gamma}b_{in}(t), \nonumber \\
& & \\
\dot{b}^\dagger(t) & = & -\frac{\gamma}{2}b^\dagger(t) +2i\chi^* b(t) +iga(t) +\sqrt{\gamma}b^\dagger_{in}(t), \nonumber \\
& &  \label{eq:qle8}
\end{eqnarray}
\end{subequations}
where the effective coupling is $g = \left| \kappa \left\langle a^b_0(t\rightarrow \infty) \right\rangle \right|$. The system is now described by the effective Hamiltonian
\begin{equation}
H_{b} = \hbar \left(\chi^* b^2 + \chi b^{\dagger 2} \right) + \hbar g \left(a b + a^\dagger b^\dagger \right). \label{blueeff}
\end{equation}
The coupling describes a non-degenerate parametric amplifier (NDPA) between the nanoresonator and cavity modes. The underlying physical process involves a drive photon being Raman scattered into a photon at the cavity resonance and a phonon at the mechanical resonance, or vice versa. The NDPA alone would result in two-mode squeezing; that is, a particular combination of quadratures of the nanoresonator and cavity will be squeezed \cite{Vitali4}. However, the NDPA puts each mode individually into a thermal state.  

The system (\ref{eq:qle5})-(\ref{eq:qle8}) is linear and homogeneous; thus, assuming stability, the steady-state is a Gaussian state of zero amplitude with fluctuations fully characterized by its correlation matrix. The system is stable provided that 
\begin{equation}
\left| \chi \right|< -\frac{g^2}{\mu} + \frac{\gamma }{4}. \label{stable3}
\end{equation}
Assuming that this is satisfied, we may Fourier transform the system to obtain
\begin{equation}
-\textbf{D}
\left[ \begin{array}{c}
a_{in}(\omega) \\
a^\dagger_{in}(-\omega) \\
b_{in}(\omega) \\
b^\dagger_{in}(-\omega)
\end{array} \right]
= \textbf{A}_b
\left[ \begin{array}{c}
a(\omega) \\
a^\dagger(-\omega) \\
b(\omega) \\
b^\dagger(-\omega)
\end{array} \right], \label{eq:inputintra}
\end{equation}
where $\textbf{D}$ denotes the damping matrix
\begin{equation}
\textbf{D}=
\left[ \begin{array}{cccc}
\sqrt{\mu} & 0 & 0 & 0 \\
0 & \sqrt{\mu} & 0 & 0 \\
0 & 0 & \sqrt{\gamma} & 0 \\
0 & 0 & 0 & \sqrt{\gamma} 
\end{array} \right],
\end{equation}
and the dynamical matrix in the frequency domain is 
\begin{equation}
\textbf{A}_b=\left[ \begin{array}{cccc}
i\omega - \frac{\mu}{2} & 0 & 0 & - ig  \\
0 & i\omega - \frac{\mu}{2} &  ig  & 0 \\
0 &- ig & i\omega -\frac{\gamma}{2} &  -2i\chi \\
 ig & 0 & 2i\chi^* & i\omega - \frac{\gamma}{2}  
\end{array} \right].
\end{equation}
Henceforth, the column vectors in (\ref{eq:inputintra}) shall be denoted by $\textbf{a}^b_{in} (\omega)$ and $\textbf{a}^b(\omega)$, respectively. Further, the output operators may be calculated in terms on the input operators using the boundary condition \cite{Gardiner9}
\begin{equation}
\textbf{a}^b _{o} (\omega) = \textbf{Da}^b(\omega)-\textbf{a}^b_{in} (\omega) = - \left ( \textbf{D}\textbf{A}^{-1}_b\textbf{D} + \textbf{1} \right ) \textbf{a}^b_{in}(\omega). \label{BC}
\end{equation}
To incorporate the effects of internal losses in the cavity, the total damping due to both internal losses and out-coupling of the field would be included in (\ref{eq:qle5})-(\ref{eq:qle8}), but only the component due to out-coupling of the field would be included in the boundary condition (\ref{BC}). This would lead to a slight reduction in the magnitude of the squeezing attainable. 

\subsubsection{Cavity Driven on Red Sideband}
Now suppose the cavity is driven on its first red sideband,
\begin{equation}
\omega_d = \omega_c - \nu , \ \ \ \ \ \ \ \ \left(i.e. \ \ \  \delta_1 = +\nu \right).
\end{equation}
Now the oscillation of the red sideband of the driving field is off-resonance and accordingly we neglect $a_-(t)$. Again assuming $\kappa << \mu , \left| \delta_1 \right|,\left| \mathcal{E}_1 \right| $ and $\mathcal{E}_1 $ real and positive, we solve (\ref{eq:kappagone}) for the steady-state amplitude at the red sideband drive frequency, $ \left\langle a^r_o (t \rightarrow \infty ) \right\rangle =-\mathcal{E}_1/\nu$. From (\ref{eq:qle2}) and (\ref{eq:qle4}), with the corresponding Hermitian conjugate equations and again dropping sideband subscripts,
\begin{subequations}
\begin{eqnarray}
\dot{a}(t) & = & -\frac{\mu}{2}a(t) +igb(t) + \sqrt{\mu}a_{in}(t), \label{eq:qle9}\\
\dot{a}^\dagger(t) & = & -\frac{\mu}{2}a^\dagger(t) -igb^\dagger(t) + \sqrt{\mu}a^\dagger_{in}(t), \\
\dot{b}(t) & = & -\frac{\gamma}{2}b(t) -2i\chi b^\dagger(t) +iga(t) + \sqrt{\gamma}b_{in}(t), \\
\dot{b}^\dagger(t) & = & -\frac{\gamma}{2}b^\dagger(t) +2i\chi^* b(t) -iga^\dagger(t) +\sqrt{\gamma}b^\dagger_{in}(t), \nonumber \\ & &  \label{eq:qle12}
\end{eqnarray}
\end{subequations}
where, equivalently to above, $g= - \kappa \left\langle a^r_o (t \rightarrow \infty ) \right\rangle $. The effective Hamiltonian is 
\begin{equation}
H_{r} = \hbar \left(\chi^* b^2 + \chi b^{\dagger 2} \right) + \hbar g \left(a^\dagger b + a b^\dagger \right). \label{redeff}
\end{equation}
The second term represents a beamsplitter-like coupling between the nanoresonator and the cavity. The underlying mechanism it describes is a Raman scattering process in which an injected drive photon and a phonon emitted from the nanoresonator result in a photon at the cavity resonance, or vice versa. 

The stability conditions are now
\begin{equation}
\left| \chi \right|< \frac{g^2}{\mu} + \frac{\gamma }{4}, \ \ \ \ \ \left| \chi \right| < \frac{\gamma + \mu}{4}, \label{stable1}
\end{equation}
and the same comments apply for the resulting steady-state as in the preceding section. Again we have 
\begin{equation}
\textbf{a}^r _{o} (\omega) = \textbf{Da}^r(\omega)-\textbf{a}^r_{in} (\omega) = - \left ( \textbf{D}\textbf{A}^{-1}_r\textbf{D} + \textbf{1} \right ) \textbf{a}^r_{in}(\omega), \label{eq:BC2}
\end{equation}
where now
\begin{equation}
\textbf{A}_r=
\left[ \begin{array}{cccc}
i\omega - \frac{\mu}{2} & 0 & ig & 0  \\
0 & i\omega - \frac{\mu}{2} &  0  & -ig \\
ig & 0 & i\omega -\frac{\gamma}{2} &  -2i\chi \\
 0 & -ig & 2i\chi^* & i\omega - \frac{\gamma}{2}  
\end{array} \right].
\end{equation}

\subsection{Cavity Driven on Two Sidebands: Blue and Red}
For the two sideband drive case, the Hamiltonian (\ref{Hint2}) leads to  
\begin{subequations}
\begin{eqnarray}
\dot{a}(t) & = & -i\mathcal{E}_1 e^{i\delta_1 t} - i\mathcal{E}_2 e^{i\delta_2 t} - \frac{\mu}{2}a(t) + \sqrt{\mu}a_{in}(t) \nonumber \\
& & \ \ -i\kappa \left[ b(t)e^{-i\nu t} + b^\dagger (t)e^{i\nu t} \right]a(t), \label{2wholesystem1} \\
\dot{b}(t) & = & -2 i \chi b^\dagger (t) - i \kappa a^\dagger (t)a(t) e^{i\nu t} - \frac{\gamma}{2}b(t) + \sqrt{\gamma} b_{in }(t), \label{2wholesystem2} \nonumber \\ & & \ \ \ \ 
\end{eqnarray}
\end{subequations}
with the input noise correlation functions (\ref{cavnoise})-(\ref{nrnoise}). Assuming that both drives are such that we are in the resolved sideband regime, that is,
\begin{equation}
\left|\delta_1 \right| , \left|\delta_2 \right|>> \mu , \label{rsb2}
\end{equation}
the same ansatz (\ref{eq:assumedalpha})-(\ref{eq:assumedbeta}) should solve (\ref{2wholesystem1})-(\ref{2wholesystem2}). Substituting, equating frequency components and also assuming $\delta_1 = -\nu $ and $\delta_2 = +\nu $ (corresponding to driving on both the red and blue sidebands), we have
\begin{widetext}
\begin{subequations}
\begin{eqnarray}
\dot{a}_0 (t) & = & -i\kappa \left[ b(t)a_-(t) + b^\dagger (t) a_+(t) \right] - \frac{\mu}{2}a_0 (t) + \sqrt{\mu}a_{0,in}(t), \label{2sb0} \\
\dot{a}_+ (t) & = & -i\kappa b(t) a_0(t) - i\mathcal{E}_1 + \left(i\nu - \frac{\mu}{2}\right)a_+(t) + \sqrt{\mu}a_{+,in}(t), \\
\dot{a}_- (t) & = & -i\kappa b^\dagger (t) a_0(t) -i\mathcal{E}_2 - \left( i\nu + \frac{\mu}{2} \right) a_-(t) + \sqrt{\mu}a_{-,in}(t), \\
\dot{b}_0 (t) & = & -2i\chi b^\dagger (t) - i\kappa \left[ a^\dagger_0(t) a_+(t) + a^\dagger_-(t)a_0(t) \right] -\frac{\gamma}{2}b(t) + \sqrt{\gamma}b_{in}(t). \label{2sbb}
\end{eqnarray}
\end{subequations}
\end{widetext}
Now setting, without loss of generality, $\mathcal{E}_1 = \mathcal{E}e^{-i\psi}$ and $\mathcal{E}_2 = -\mathcal{E}e^{i\psi}$ where $\mathcal{E} $ is real, and assuming $\kappa << \mu , \mathcal{E}$, we have the steady-state amplitudes at the drive frequencies $\left\langle a^{br}_+ (t \rightarrow \infty ) \right\rangle = \mathcal{E}e^{i\psi}/\nu $ and $ \left\langle a^{br}_- (t \rightarrow \infty ) \right\rangle = \mathcal{E}e^{-i\psi}/\nu $. The introduced $\psi$ describes the relative phase between the two cavity drives. Then (\ref{2sb0}) and (\ref{2sbb}), with the corresponding Hermitian conjugate equations and again dropping sideband subscripts, lead to
\begin{subequations}
\begin{eqnarray}
\dot{a}(t) & = & -ig \left[b(t)e^{-i\psi}+b^\dagger(t) e^{i\psi} \right] -\frac{\mu}{2}a(t) + \sqrt{\mu}a_{in}(t), \nonumber \\ \label{eq:qle13} \\
\dot{a}^\dagger(t) & = & ig \left[b(t)e^{-i\psi}+b^\dagger(t) e^{i\psi} \right] -\frac{\mu}{2}a^\dagger(t) + \sqrt{\mu}a^\dagger_{in}(t), \nonumber \\ \label{eq:qle14} 
\end{eqnarray}
\begin{eqnarray}
\dot{b}(t) & = & -2i\chi b^\dagger(t) -ige^{i\psi} \left[a(t)+a^\dagger(t) \right] \nonumber \\ & & \ \ \ \  -\frac{\gamma}{2}b(t) + \sqrt{\gamma}b_{in}(t), \label{eq:qle15}  \\
\dot{b}^\dagger(t) & = & 2i\chi^* b(t) + ige^{-i\psi} \left[a(t)+a^\dagger (t)\right] \nonumber \\ & & \ \ \ \ - \frac{\gamma}{2}b^\dagger(t) + \sqrt{\gamma}b^\dagger_{in}(t), \label{eq:qle16}
\end{eqnarray}
\end{subequations}
where, equivalently to above, the effective coupling is $g = \kappa \left|\left\langle a^{br}_+ (t\rightarrow \infty ) \right\rangle\right| =  \kappa \left|\left\langle a^{br}_- (t\rightarrow \infty ) \right\rangle \right| $. The system remains stable provided that
\begin{equation}
\chi < \frac{\gamma}{4}. \label{stable2}
\end{equation}
Note that this stability threshold is more stringent than that (\ref{stable3}) for the red sideband drive, but less stringent than that (\ref{stable2}) for the blue sideband drive. Now the system dynamics are governed by the effective Hamiltonian,
\begin{equation}
H_{br} = \hbar \left( \chi^* b^2 + \chi b^{\dagger 2} \right) + \hbar g \left( a + a^\dagger \right) \left( be^{-i\psi} + b^\dagger e^{i\psi} \right). \label{rbeff}
\end{equation}
The second term has the form of a back-action evading measurement of a quadrature of the nanoresonator motion; which quadrature is measured depends on the relative phase of the two cavity drives. Physically, the Raman processes corresponding to the injection of a photon at the cavity resonance and the absorption or emission of a phonon by the nanoresonator are both possible and occur at the same rate. 

Assuming stability, we may Fourier transform (\ref{eq:qle13})-(\ref{eq:qle16}) and apply the usual boundary conditions to find 
\begin{equation}
\textbf{a}^{br} _{out} (\omega) = \textbf{Da}^{br}(\omega)-\textbf{a}^{br}_{in} (\omega) = - \left ( \textbf{D}\textbf{A}^{-1}_{br}\textbf{D} + \textbf{1} \right ) \textbf{a}^{br}_{in}(\omega), \label{eq:BC3}
\end{equation}
where the dynamical matrix is now
\begin{equation}
\textbf{A}_{br}=
\left[ \begin{array}{cccc}
i\omega - \frac{\mu}{2} & 0 & -ige^{-i\psi} & -ige^{i\psi}  \\
0 & i\omega - \frac{\mu}{2} &  ige^{-i\psi}  & ige^{i\psi} \\
-ige^{i\psi} & -ige^{i\psi} & i\omega -\frac{\gamma}{2} &  -2i\chi \\
ige^{-i\psi} & ige^{-i\psi} & 2i\chi^* & i\omega - \frac{\gamma}{2}  
\end{array} \right].
\end{equation}

\section{SQUEEZING OF QUADRATURES OF NANORESONATOR MOTION }
To observe squeezing, we are interested in the quadratures of the nanoresonator motion,
\begin{subequations}
\begin{eqnarray}
X'_m & = & be^{-i\phi} + b^\dagger e^{i\phi} , \label{mquaddef1} \\
Y'_m & = & -i (be^{-i\phi} - b^\dagger e^{i\phi}) , \label{mquaddef2}
\end{eqnarray}
\end{subequations}
where $\phi $ is the rotation angle relative to the conventional position and momentum quadratures. The quadrature normally-ordered variances (that is, squeezing) are then
\begin{subequations}
\begin{eqnarray}
S_{X'_m} & = & \left\langle  : X'_m , X'_m : \right\rangle \nonumber \\ & = & e^{-2i\phi}\left\langle b^2 \right\rangle + e^{2i\phi} \left\langle b^{\dagger 2} \right\rangle + 2\left\langle b^\dagger b \right\rangle , \label{nrsqueezeX} \\
S_{Y'_m} & = & \left\langle  : Y'_m , Y'_m : \right\rangle \nonumber \\ & = & -e^{-2i\phi}\left\langle b^2 \right\rangle - e^{2i\phi} \left\langle b^{\dagger 2} \right\rangle + 2\left\langle b^\dagger b \right\rangle . \label{nrsqueezeY}
\end{eqnarray}
\end{subequations}
These may be calculated by writing quantum Langevin equations for all second moments of the nanoresonator and cavity operators, and solving for their expectations in the steady-state. 

With $\chi$ real, the optimally squeezed quadrature is $Y'_m $ with $\phi = -\pi /4 $, irrespective of the driving conditions, provided that we set $\psi = \pi /4$ for the two sideband drive case. Any quadrature may be optimally squeezed through suitable choice of the phase of the parametric driving; with $Arg\left[ \chi \right] = -\pi /2 \ \ \ (+\pi /2 )$ the position (momentum) quadrature is squeezed. Below we quote results for $\chi $ real and the squeezed $(Y'_m)$ quadrature; the same expressions apply to the optimally squeezed quadrature when a non-zero parametric driving phase is adopted. We find, for driving on the blue, red, and blue and red sidebands, 
\begin{subequations}
\begin{eqnarray}
S^b_{Y'_m} & = &  2 \left[ \mu (n^0_m\gamma - 2\chi) (\gamma + \mu + 4\chi ) \right. \nonumber \\ & & \ \ \ \left. -4g^2 (n^0_m\gamma -\mu -2\chi ) \right] / \nonumber \\ & & \ \ \ \ \ \  \left[(\gamma + \mu + 4\chi) (\mu \gamma + 4\mu \chi - 4g^2 )\right] , \nonumber \\ & & \\
S^r_{Y'_m} & = & \frac{2(n^0_m\gamma - 2\chi )(4g^2 + \mu \gamma + \mu^2 +4\mu\chi ) }{\left( \gamma + \mu + 4\chi \right)\left( 4g^2 + \mu \gamma + 4\mu \chi \right) }, \\
S^{br}_{Y'_m} & = & \frac{2n^0_m\gamma - 4\chi }{\gamma + 4\chi }.
\end{eqnarray}
\end{subequations}
At the threshold (\ref{stable1}), assuming $4g^2 < \mu^2$, for the red sideband drive, 
\begin{eqnarray}
S^r_{Y'_m} & = & -\frac{1}{2}\frac{8g^2 + 2\gamma\mu + \mu^2 }{4g^2 + 2\gamma\mu + \mu^2 } \nonumber \\ & & \ \ \  + n^0_m \frac{\gamma \mu (8g^2 + 2\gamma\mu + \mu^2)}{(4g^2 + \gamma \mu)(4g^2 + 2\gamma\mu + \mu^2 )}.
\end{eqnarray}
For all driving conditions, at threshold and in the adiabatic limit, $S_{Y'_m} \rightarrow -\frac{1}{2}+n^0_m$ \cite{Milburn5,Collett6} and the noise in the conjugate quadrature $(X'_m)$ diverges.  

The squeezing of a quadrature of the cavity field $(S_{X'_c}$ or $S_{Y'_c})$ is given by (\ref{nrsqueezeX}) or (\ref{nrsqueezeY}) with the replacement $b \rightarrow a $. Here we quote the results for the red sideband drive as these shall be useful later,
\begin{equation}
S^r_{Y'_m} = S^r_{X'_c} + \frac{ 2\mu (n^0_m\gamma - 2\chi ) }{4g^2 + \gamma \mu + 4\mu \chi }, \label{nrcav}
\end{equation}
such that at the threshold (\ref{stable1}), assuming $4g^2<\mu^2$,
\begin{equation}
S^r_{Y'_m} = S^r_{X'_c} - \frac{1}{2} + \frac{\gamma \mu}{4g^2 + \gamma \mu}n^0_m .
\end{equation}
Thus squeezing of the internal cavity field implies squeezing of a nanoresonator quadrature provided that
\begin{equation}
n^0_m < \frac{4g^2 + \gamma \mu}{2\gamma \mu}. \label{mechanicalbathcondition}
\end{equation}

\section{QUANTUM-LIMITED TRANSDUCER OF NANORESONATOR MOTION}
\subsection{Adiabatic Elimination}
The quantum Langevin equations (\ref{eq:qle5})-(\ref{eq:qle8}), (\ref{eq:qle9})-(\ref{eq:qle12}) and (\ref{eq:qle13})-(\ref{eq:qle16}) may each be mapped back onto an effective quantum optics master equation, of the form
\begin{equation}
\dot{\rho} = -\frac{i}{\hbar}\left[ H_e, \rho \right] + \gamma (n^0_m + 1 )\mathcal{D}\left[b \right]\rho + \gamma n^0_m \mathcal{D}\left[b^\dagger \right] \rho + \mu \mathcal{D}\left[a \right]\rho , \label{effme}
\end{equation}
where $H_e $ denotes the effective Hamiltonian, (\ref{blueeff}), (\ref{redeff}) or (\ref{rbeff}), and the superoperator $\mathcal{D}$ is defined via its action,
\begin{equation}
\mathcal{D}\left[s\right]\rho = s\rho s^\dagger - \frac{1}{2}s^\dagger s\rho - \frac{1}{2}\rho s^\dagger s.
\end{equation}
 
We assume the cavity is heavily damped such that the cavity mode at the sideband of the driving field will have few photons and it will be slaved to the nanoresonator mode. This is the adiabatic limit, 
\begin{equation}
\left| \frac{\left\langle H_e \right\rangle }{\mu} \right| \approx {\rm max} \left[ \frac{g}{\mu}, \frac{\chi}{\mu} \right] = \epsilon << 1, \label{adiabaticlimit}
\end{equation}
and accordingly we shall now adiabatically eliminate the cavity mode \cite{Wiseman2}. Expanding the nanoresonator-cavity density operator in powers of $\epsilon$ using a low photon number basis for the cavity, 
\begin{eqnarray}
\rho (t) & = & \rho_{00}(t)\otimes \left|0 \right\rangle_a\left\langle 0\right| + \left[ \rho_{01}(t)\otimes \left|0 \right\rangle_a\left\langle 1\right| + H.c. \right] \nonumber \\ & & \ \ \  + \rho_{11}(t)\otimes \left|1\right\rangle_a\left\langle 1\right|  \nonumber \\ & & \ \ \ \ \ + \left[ \rho_{02}(t)\otimes \left|0 \right\rangle_a\left\langle 2\right| + H.c. \right] + \mathcal{O}(\epsilon^3),
\end{eqnarray}
and substituting this, neglecting the last two second-order terms, into (\ref{effme}), one obtains a closed system for the density operators $\rho_{00}$, $\rho_{01}$, $\rho_{10}$ and $\rho_{11}$. Assuming the off-diagonal elements are most rapidly damped, we may solve for their steady-states. Then, defining the nanoresonator density operator,
\begin{equation}
\rho_m (t) = {\rm Tr_c} \ \rho (t) = \rho_{00}(t) + \rho_{11}(t) ,
\end{equation}
where ${\rm Tr_c}$ denotes the trace over the cavity mode, we obtain the master equations 
\begin{subequations}
\begin{eqnarray}
\dot{\rho}^b_m & = & -i \left[\chi^* b^2 + \chi b^{\dagger 2}, \rho^b_m \right] + \frac{4g^2}{\mu} \mathcal{D}\left[ b^\dagger \right]\rho^b_m \nonumber \\ & & \ \ \ \ \ \ + \gamma (n^0_m + 1) \mathcal{D}\left[b \right] + \gamma n^0_m \mathcal{D}\left[ b^\dagger \right] \rho^b_m , \label{MEb} \\
\dot{\rho}^r_m & = & -i \left[\chi^* b^2 + \chi b^{\dagger 2}, \rho^r_m \right] + \frac{4g^2}{\mu} \mathcal{D}\left[b \right]\rho^r_m \nonumber \\ & & \ \ \ \ \ \ + \gamma (n^0_m +1 )\mathcal{D}\left[b \right]\rho^r_m + \gamma n^0_m \mathcal{D}\left[b^\dagger \right]\rho^r_m , \nonumber \\ & & \label{MEr} \\
\dot{\rho}^{br}_m & = & -i \left[ \chi^*b^2 + \chi b^{\dagger 2}, \rho^{br}_m \right] \nonumber \\ & & \ \ \ + 2\frac{4g^2}{\mu} \mathcal{D}\left[be^{-i\psi} + b^\dagger e^{i\psi} \right] \rho^{br}_m  \nonumber \\ & & \ \ \ \ \ \ + \gamma (n^0_m + 1) \mathcal{D}\left[b \right]\rho^{br}_m + \gamma n^0_m \mathcal{D}\left[b^\dagger \right]\rho^{br}_m . \nonumber \\ \label{MEbr}
\end{eqnarray}
\end{subequations}
In all cases, the nanoresonator is damped both into its mechanical bath and by virtue of its coupling to the microwave cavity. For the blue sideband drive a heating term appears in (\ref{MEb}) and for the red sideband drive a cooling term appears in (\ref{MEr}). The second term in (\ref{MEbr}) describes diffusion in the quadrature conjugate to that which is measured. No noise is added to the quadrature that is measured, and thus this arrangement is preferable for the measurement of a quadrature \cite{Clerk}. An optimal weak, continuous measurement of the quadrature may be realized using the prescription of Clerk \cite{Clerk4}. 

The final mean thermal phonon number of the nanoresonator, for each driving condition, follows from (\ref{MEb})-(\ref{MEbr}). For the blue sideband drive it will be heated according to $n^b_m = (\Gamma + \gamma n^0_m)/(\gamma - \Gamma)$; for the blue and red sideband drive it will be heated according to $n^{br}_m = n^0_m + 2\Gamma / \gamma $; and for the red sideband drive it will be cooled according to 
\begin{equation}
n^r_m = \frac{\gamma n^0_m }{\Gamma + \gamma}, \label{finalphononsred}
\end{equation}
where $ \Gamma = 4g^2/\mu$. This is the result of Marquardt \emph{et al.} \cite{Girvin1} in the extreme resolved sideband limit, $\left( \mu / 4\nu \right)^2 \rightarrow 0 $. To reproduce their full result within a master equation formalism, one must retain the neglected off-resonant sideband of the driving field, and adiabatically eliminate the cavity modes using a projection operator approach \cite{Breuer}. 

A quantum-limited weak, continuous measurement is one for which the noise added to the signal is determined only by the measurement back-action noise \cite{Walls-GJM}. In a complementary way, this means that the only noise added to the system comes exclusively from the measurement process itself. In a quantum-limited measurement, the observer may gain from the environment all the information needed to completely describe the state of the measured system at any time. This requires that the only dissipative channel in the master equation for the system corresponds to the output channel that is monitored. With adiabatic elimination, the cavity field acts like a dissipative channel for the mechanical resonator while simultaneously providing the channel by which the measurement is made. We effectively have a quantum-limited measurement provided that 
\begin{equation}
n^{b/r/br}_m<<1 , \ \ \ \  \gamma << 4g^2/\mu . \label{qltransducer}
\end{equation} 
These conditions can only be satisfied, without excessive constraints on the other parameters, by driving the cavity on the red sideband alone. Results for all three cases shall be discussed below, though the focus shall be on the case where the cavity is driven on its red sideband only. 


\subsection{Nanoresonator and Cavity Quadratures}
The directly measurable quantities are the frequency components of the output microwave field quadratures, 
\begin{subequations}
\begin{eqnarray}
X'_c(\omega ) & = & a(\omega )e^{-i\theta } + a^\dagger (\omega ) e^{i\theta} , \label{quaddef1} \\
Y'_c (\omega ) & = & - i \left[ a (\omega )e^{-i\theta} - a^\dagger (\omega )e^{i\theta} \right] , \label{quaddef2}
\end{eqnarray}
\end{subequations}
where $\theta$ is the local oscillator phase in our homodyne detection scheme.

Under the condition (\ref{adiabaticlimit}) we may write formal solutions to the quantum Langevin equations (\ref{eq:qle5}), (\ref{eq:qle9}) and (\ref{eq:qle13}), take their Fourier transforms and apply the usual boundary condition to find
\begin{subequations}
\begin{eqnarray}
a^b_o (\omega ) & = & -\frac{2ig}{\sqrt{\mu}}b^\dagger (-\omega ) + a_{in}(\omega ), \\
a^r_o (\omega ) & = & \frac{2ig}{\sqrt{\mu}}b (\omega ) +a_{in}(\omega ), \\
a^{br}_o (\omega ) & = & -\frac{2ig}{\sqrt{\mu}} \left[ b(\omega )e^{-i\psi} + b^\dagger (-\omega )e^{i\psi} \right] + a_{in}(\omega ). \nonumber \\
& & 
\end{eqnarray}
\end{subequations}
Thus, by appropriate choice of the local oscillator phase (and the relative phase of the cavity drives in the two sideband case), one can monitor any quadrature of the nanoresonator motion via the output microwave field. 

\section{OUTPUT MICROWAVE FIELD QUADRATURE SQUEEZING SPECTRA}

\begin{figure*}[th!]
\includegraphics[scale=0.5]{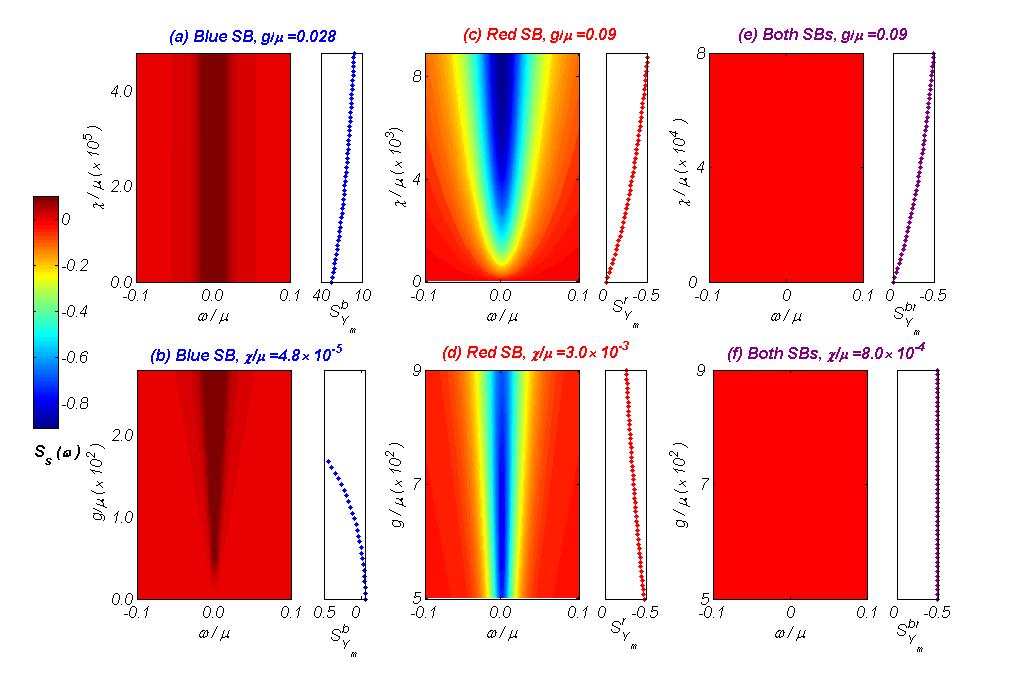}
\caption{Zero-temperature squeezing spectra $S_s\left( \omega \right) $ and nanoresonator quadrature squeezing $S_{Y'_m}$ for driving the cavity (a) on the blue sideband as a function of $\chi / \mu$ with $g/\mu = 0.028$, (b) on the blue sideband as a function of $g/\mu$ with $\chi / \mu = 4.8\times 10^{-5} $, (c) on the red sideband as a function of $\chi / \mu $ with $g/\mu = 0.09$, (d) on the red sideband as a function of $g/\mu $ with $\chi / \mu = 0.003 $, (e) on both the blue and red sidebands as a function of $\chi / \mu$ with $g / \mu = 0.09$, and (f) on both the blue and red sidebands as a function of $g/\mu $ with $\chi / \mu = 8.0\times 10^{-4}$. Parameters are chosen such that the system remains below threshold. For all plots, $\omega_c = 2\pi \times 6\, {\rm GHz} $, $\nu = 2\pi \times 20\, {\rm MHz} $, $m=10^{-15}\, {\rm kg} $, $\gamma = 1.26\times 10^3\, {\rm s^{-1}}$, and $\mu = 3.77 \times 10^5\, {\rm s^{-1}}$. The maximal squeezing of a nanoresonator quadrature attainable is $-3{\rm dB}$, achieved on threshold for the red sideband drive and for the blue and red sideband drive cases. Excess noise is introduced in the blue sideband drive case due to the effective NDPA nanoresonator-cavity coupling. The quantum non-demolition interaction in the two sideband drive case means that the microwave field cannot be squeezed unconditionally. For the red sideband drive case, squeezing of the microwave field implies squeezing of a nanoresonator quadrature. Further, approaching threshold, near-maximal squeezing of the microwave field is achievable for experimentally feasible parameters. }
\label{fig-10}
\end{figure*}

\subsection{Calculation}
The output microwave field quadrature squeezing spectra are given by the normally-ordered variances,
\begin{subequations}
\begin{eqnarray}
S_{X'_c} (\omega)& = & \left\langle :X'_c (\omega), X'_c (\omega) :\right\rangle \nonumber \\
& = & e^{-2i\theta} \left\langle a_o(\omega),a_o(\omega)\right\rangle + e^{2i\theta} \left\langle a^\dagger_o(\omega),a^\dagger_o(\omega)\right\rangle \nonumber \\ 
& & \ \ + 2\left\langle a^\dagger_o(\omega),a_o(\omega)\right\rangle , \label{squeeze1} \\
S_{Y'_c} (\omega)& = & \left\langle :Y'_c (\omega), Y'_c (\omega) :\right\rangle \nonumber \\
& = & -e^{-2i\theta} \left\langle a_o(\omega),a_o(\omega)\right\rangle - e^{2i\theta} \left\langle a^\dagger_o(\omega),a^\dagger_o(\omega)\right\rangle \nonumber \\ 
& & \ \ + 2\left\langle a^\dagger_o(\omega),a_o(\omega)\right\rangle . \label{squeeze2} 
\end{eqnarray}
\end{subequations}
Each of the variances on the right hand sides of (\ref{squeeze1}) and (\ref{squeeze2}) may be expressed in terms of the variances of input operators using (\ref{BC}), (\ref{eq:BC2}) or (\ref{eq:BC3}). These spectra are related to the nanoresonator quadrature squeezing through (\ref{nrcav}) and
\begin{equation}
S_{X'_c} = \mu \int^{+\infty}_{-\infty} \left\langle : X'_c(\omega ), X'_c(\omega ):\right\rangle d\omega .
\end{equation}
Thus the observation of quantum squeezing in any frequency component of the microwave field implies squeezing of the internal cavity field, and for the red sideband drive, through (\ref{nrcav}), squeezing of a nanoresonator quadrature provided that (\ref{mechanicalbathcondition}) is satisfied.

The output microwave field must be amplified for detection. Now a linear, phase-preserving amplifier with gain $A \geq 2 $ will destroy quantum squeezing \cite{Caves1}. However, it may still be possible to infer quantum squeezing by comparing the amplified noise level of the parametrically deamplified quadrature with and without the pump applied. The amplified signal is given by
\begin{equation}
c(\omega) = \sqrt{A}a_{o}(\omega)+ \sqrt{A-1}d^\dagger (\omega), \label{signalout}
\end{equation}
where $d(\omega)$, with the non-zero correlation function
\begin{equation}
\left\langle d^\dagger (\omega), d(\omega ') \right\rangle = \delta (\omega - \omega ').
\end{equation}
specifies the noise at the auxiliary amplifier input. The effect of amplifier gain $A$ and amplifier noise $n_a$ is to add a frequency-independent noise floor, 
\begin{equation}
S^{A}_s (\omega ) = AS^b_s (\omega ) + 2(A-1)(n_a + 1) .
\end{equation}

\subsection{Experimentally Feasible Parameters}
We now estimate experimentally accessible parameters. The cavity resonance frequency will be $\omega_c/2\pi=6 \, {\rm GHz}$ \cite{Frunzio} and the nanoresonator frequency will be $\nu/2\pi=20 \, {\rm MHz}$ \cite{Schwab1}. The cavity impedance is $50\, \Omega $, such that the cavity is described by the equivalent inductance $L = 1.33\, {\rm nH} $ and the equivalent capacitance is $C = 0.531 \, {\rm pF} $. Assuming a nanoresonator mass of $10^{-15}\, {\rm kg}$, the ground state uncertainty in nanoresonator position is $\Delta x = 20.5\, {\rm fm}$. Approximating $d=80\, {\rm nm}$ and $\beta =0.002$, then $\kappa = 9.6\, {\rm s^{-1}}$. Microwave cavities can be fabricated with $Q_c = 10^5$ and nanoresonators with $Q_m = 10^5$, with corresponding damping rates are $\mu = 3.77\times 10^5\, {\rm s^{-1}}$ and $\gamma = 1.26\times10^3\, {\rm s^{-1}}$. 

The fiducial coupling $g/\mu = 0.09 $ corresponds to $\mathcal{E} = 4.441 \times 10^{11}\, {\rm s^{-1}}$, and a photon number at the drive frequency of $n_d = 1.249\times 10^7 $, a peak drive voltage of $13.7\, {\rm mV}$, and a circulating power of $1.87\, {\rm \mu W}$, below the typical critical circulating powers \cite{Lehnert3} at which the cavity response becomes nonlinear \cite{Dahm1}. The fiducial parametric driving strength $\left|\chi \right|/ \mu =0.01 $ corresponds to $k_0 = 3.79\times 10^{-3}\, {\rm kgs^{-2}} $, or a $0.024\% $ change in the unperturbed effective spring constant. Assuming $x_{c0} = 80\, {\rm nm} $ and $C_{c0}=200\, {\rm aF} $, the required adjustment is easily achieved by $V_0V_P =0.121\, {\rm V^2} $. Further, $\Gamma = 5.49 \times 10^5\, {\rm s^{-1}} $ such that ground-state cooling should be feasible. The conditions (\ref{rsb1}) or (\ref{rsb2}), (\ref{stable1}) or (\ref{stable2}), and (\ref{qltransducer}), can then, at least in principle, be easily satisfied with reasonable parameters. 

\subsection{Results: Zero Temperature Bath}

Assuming a zero temperature mechanical bath $(n^0_m = 0 ) $, no amplification of the output field $(A=1) $, and that $\chi $ is real, the optimally squeezed microwave field quadrature is $X'_c(\omega )$ with $\theta = -\pi /4 $ for the blue or red sideband, and $\theta = 0$ for the blue and red sideband drive. By appropriately setting the phase of the parametric drive and the relative cavity driving phase, any quadrature of the microwave field may be optimally squeezed. Specifically, for the red sideband drive, with $\chi=\pi /2$ the amplitude quadrature of the microwave field is optimally squeezed, and for $\chi = -\pi /2$, the phase quadrature is optimally squeezed. However, these optimal squeezing spectra have the same functional form and henceforth we shall simply refer to the squeezed and anti-squeezed quadratures (subscripts s and as, respectively). The squeezing spectra are
\begin{widetext}
\begin{subequations}
\begin{eqnarray}
S^b_s (\omega ) & \equiv &  S^b_{X'_c} (\omega) = \frac{32g^2\mu (\gamma - 2\chi)}{ (4g^2 -\gamma\mu + 4\mu\chi)^2 + 4\left[8g^2 + \mu^2 + (\gamma - 4\chi)^2\right]\omega^2 + 16\omega^4  } , \label{Sb1} \\
S^r_{s} (\omega ) & \equiv & S^r_{X'_c}(\omega ) = \frac{-64g^2\mu \chi }{(4g^2 +\gamma\mu + 4\mu\chi)^2 + 4\left[-8g^2 + \mu^2 + (\gamma + 4\chi)^2\right]\omega^2 + 16\omega^4  }, \label{Sr1} \\
S^{br}_{s}(\omega )  & \equiv & S^{br}_{X'_c} (\omega ) = 0 , \ \ \ \ \  S^{br}_{as}(\omega )  \equiv S^{br}_{Y'_c} (\omega ) = \frac{ 64g^2\gamma \mu }{ (\mu^2 + 4\omega^2 ) \left[ (\gamma + 4\chi )^2 + 4\omega^2 \right] } . \label{Sbr2}
\end{eqnarray}
\end{subequations}
\end{widetext}

These squeezing spectra are plotted in Fig.~2, along with the corresponding nanoresonator quadrature squeezing, as a function of $\chi$ and $g$. For all driving conditions, the maximum attainable squeezing of the nanoresonator quadrature is $-3dB$. For the blue sideband drive, this is achieved only for a vanishingly small coupling (b); for the red sideband drive, this is attained near threshold (d); and for the blue and red sideband drive, this is attained on threshold independently of the coupling (f). Clearly, the achievable squeezing increases with increasing parametric driving strength for $g$ fixed, and decreases with increasing coupling for $\chi$ fixed. The latter may be attributed to increased back-action noise. 

For the blue sideband drive the output microwave field cannot be squeezed, and it is easily seen that the observation of squeezing $(S^b_s (\omega ) <0 )$ is incompatible with the stability condition (\ref{stable3}). The excess noise introduced in this case is due to the effective NDPA nanoresonator-cavity coupling. For the red sideband drive, near-maximal squeezing of the output microwave field is attained with reasonable experimental parameters and with sufficient bandwidth for detection. Here, at zero temperature, squeezing of a frequency component of the microwave field necessarily implies squeezing of a nanoresonator quadrature. For driving on both sidebands, squeezing of the output microwave field, at least unconditionally, cannot be observed. The least noisy quadrature is characterized by vacuum noise at all $\chi$ and $g$.

\begin{figure}[th!]
\includegraphics[scale=0.23]{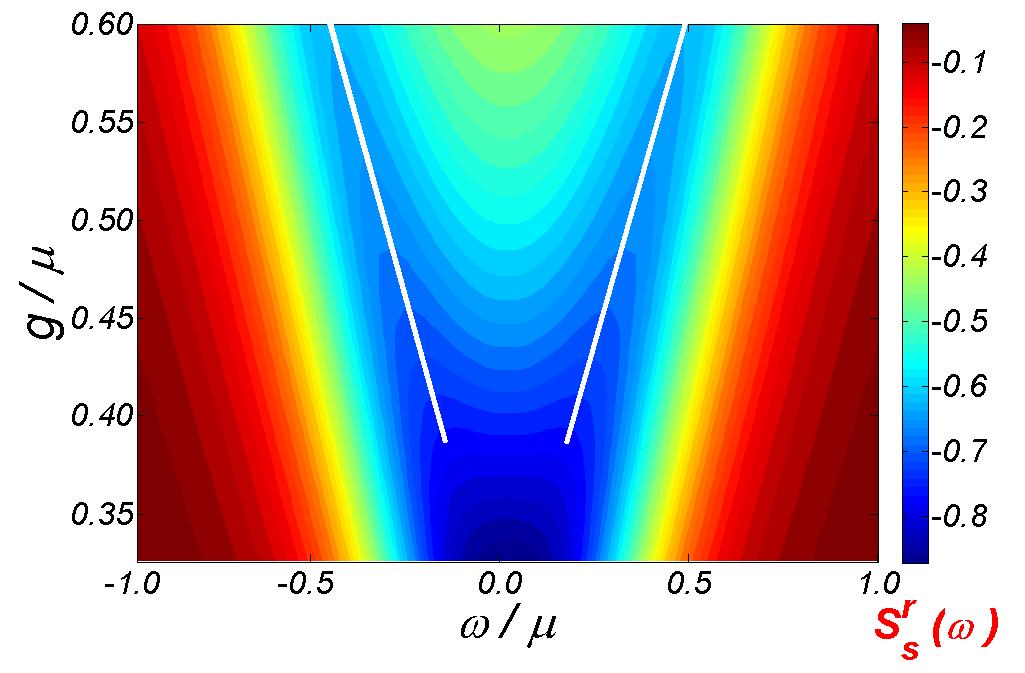}
\caption{Squeezing spectrum for driving on the red sideband with $\chi / \mu =0.1 $ and a zero-temperature mechanical bath. Normal-mode splitting, indicated by the white lines, is observed in the regime $8g^2 > \mu^2 + \left( \gamma + 4\chi \right)^2$. The normal modes asymptotically tend to the frequencies $\omega_c \pm g $. However, the required circulating power is $83\, {\rm \mu W}$, over $40$ times that presently feasible in superconducting microwave cavities. } 
\label{fig-13}
\end{figure}

Considering (\ref{Sr1}), if $ 8g^2 > \mu^2 + \left( \gamma + 4\chi \right)^2 $, then squeezing maxima would appear off-resonance, as shown in Fig.~\ref{fig-13}. This is simply classical normal-mode splitting, resulting in maximal squeezing of the cavity modes at frequencies tending to $\omega_c \pm g $ asymptotically. The requirement here is compatible with the stability conditions, though we would no longer be in the adiabatic limit. Further, with the assumed parameters, satisfying this requirement would result in a circulating power beyond that at which the cavity response becomes nonlinear. 

\begin{figure*}[th!]
\includegraphics[scale=0.5]{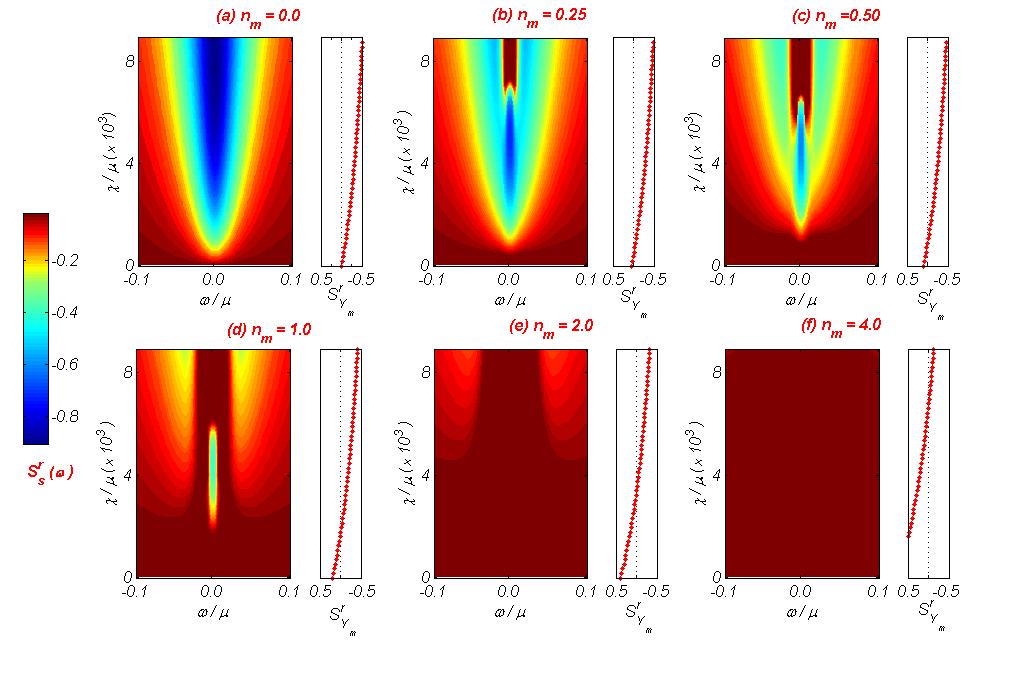}
\caption{Squeezing spectra $S^r_s (\omega )$ as filled contour plots for driving on the red sideband and mechanical bath occupation $n^0_m =$ $(a) \ \ 0.0$, $(b) \ \ 0.25$, $(c) \ \ 0.5$, $(d) \ \ 1.0$, $(e) \ \ 2.0 $ and $(f) \ \ 4.0 $, as a function of $\chi /\mu $ with $g/\mu =0.09 $. The corresponding nanoresonator quadrature squeezing is also plotted. Other parameters are $\gamma / \mu =0.003334 $ and $\mu = 3.77 \times 10^5\, {\rm s^{-1}} $. The darkest shade of red in these plots corresponds to anti-squeezing of the quadrature without indicating the magnitude of the anti-squeezing. Clearly, there is an optimal $\chi $ below threshold for a finite temperature mechanical bath. Beyond this, optimal squeezing occurs off-resonance, and the squeezing attainable is slightly reduced.  } 
\label{fig-12}
\end{figure*}

\subsection{Results: Finite Temperature Bath}
Squeezing of the output field is degraded by thermal noise on the nanoresonator. Squeezing spectra for driving on the red sideband, as a function of $\chi / \mu $, are shown in Fig.~\ref{fig-12}, along with the corresponding nanoresonator quadrature squeezing. Similar behaviour is observed for driving on the blue sideband alone, or for driving on both sidebands. 

The enhanced phase fluctuations of the nanoresonator thermal state cause the nanoresonator to couple to off-resonant components of the microwave field. Thus, above a critical parametric drive strength, optimal squeezing is obtained in the off-resonant components of the microwave field; as clearly seen in (b), (c) and (d). Excess noise is added to the microwave field on resonance beyond this critical parametric driving strength, a feature that may be attributed to reduced coupling to the squeezed quadrature. Increased thermal noise on the nanoresonator also leads to an increased threshold on the parametric driving strength (a)-(d); the parametric drive must first squeeze the classical fluctuations before reducing the quantum fluctuations below the vacuum level. Increasing further the mean phonon number of the nanoresonator thermal state, quantum squeezing is only observed off-resonance (e), and eventually, no quantum squeezing is observed at all (f). The correspondence between the squeezing of any component of the output microwave field and squeezing of a nanoresonator quadrature is observed in all the plots of Fig.~\ref{fig-12}.  

\subsection{Experimental Feasibility}
The experimental observation of quantum squeezing would proceed along similar lines to the detection of microwave squeezing in a Josephson parametric amplifier \cite{CB1}. It is assumed that the cavity itself may be treated as a parametric amplifier; a good approximation provided that $4\chi >> \gamma $. This regime can only be entered with the red sideband drive alone. One would find the total noise added by the transducer and the total gain using calibrated noise sources and the parametrically amplified quadrature, and then characterize the amplifier using the same technique but with a detuned local oscillator. Then one could confirm that the nanoresonator is in its quantum ground state. Turning to the parametrically deamplified quadrature, it should be possible to infer quantum squeezing by comparing the noise spectral density with and without the parametric drive. 

As pointed out by Di$\acute{o}$si \cite{Diosi2} and encountered in an optomechanical setting by Schliesser \emph{et al.} \cite{Kippenberg2}, the effectiveness of the resolved sideband cooling technique will be limited by phase noise on the cavity drive. This noise will also lead to a fluctuating effective coupling, and a likely degradation in the squeezing achievable. The effect of phase noise and also amplitude fluctuations on the parametric drive has been well studied for an optical DPA \cite{Zubairy1,Zubairy2}. Pump phase fluctuations lead to the large uncertainty of the amplified quadrature being mixed in with the uncertainty of the squeezed quadrature. Calculating the magnitude of these effects, and hence performing a complete evaluation of the experimental feasibility of this scheme, would require a detailed model of the microwave source. 

\section{CONCLUSIONS}
We have studied a system composed of a parametrically-driven nanomechanical resonator capacitively coupled to a microwave cavity detector. A near quantum-limited measurement of a quadrature of the nanoresonator motion may be realized by driving the cavity on its sidebands corresponding to the mechanical resonance frequency. The nanoresonator motion can be squeezed via the parametric drive, and this squeezing may be inferred from measurement of squeezing of the microwave field output from the cavity. By driving the cavity on the red sideband alone one may, in principle, simultaneously perform this measurement and cool the nanoresonator to near its quantum ground state such that quantum squeezing of the nanoresonator motion, and its detection, is feasible.

\section{ACKNOWLEDGEMENTS}

We acknowledge support from the Australian Research Council. GJM is supported by an ARC Federation Fellowship. MJW is supported by an ARC Australian Postgraduate Award. MJW thanks Christian Weedbrook for useful discussions.

\end{document}